\documentclass{emulateapj}
\usepackage{psfig}
\usepackage{subfigure}
\usepackage{lscape}

\usepackage{verbatim}

\newcommand{\etal}{{et~al.}}

\newcommand\bootes{Bo\"{o}tes}

\bibpunct[;]{(}{)}{;}{a}{}{,}
\def\gtsima{$\; \buildrel > \over \sim \;$}
\def\gsim{\lower.7ex\hbox{\gtsima}}

\usepackage[usenames]{color}

\shorttitle{The Dirt on Dry Mergers}
\shortauthors{Desai \etal}

\begin{document}

\title{The Dirt on Dry Mergers}

\author{Vandana~Desai\altaffilmark{1}, 
Arjun~Dey\altaffilmark{2},
%M.~J.~I~Brown\altaffilmark{4}, 
%Lee~Armus\altaffilmark{2},
%Matthew~L.~N.~Ashby\altaffilmark{6},
%Mark Brodwin\altaffilmark{6, 7}, 
%Shane~Bussmann\altaffilmark{8},
%Buell~T.~Jannuzi\altaffilmark{2},
%Jason~Melbourne\altaffilmark{9},
Emma~Cohen\altaffilmark{3},
Emeric~Le~Floc'h\altaffilmark{4},
B.~T.~Soifer\altaffilmark{1,3}
%Dan~Stern\altaffilmark{10},
%Steve~P.~Willner\altaffilmark{6},
%AGES folks since we use their redshifts} 
}

%\altaffiltext{1}{Based on observations made with the \textit{Spitzer
%Space Telescope}, operated by the Jet Propulsion Laboratory under NASA
%contract 1407.}

\altaffiltext{1}{Spitzer Science Center, California Institute of
Technology, Pasadena CA 91125; desai@ipac.caltech.edu}

\altaffiltext{2}{National Optical Astronomy Observatory, Tucson AZ
85726-6732}

\altaffiltext{3}{California Institute of Technology, 1200 East California Blvd, Pasadena CA 91125}

%\altaffiltext{4}{Spitzer Fellow; Institute for Astronomy, University of Hawaii,
%Honolulu HI 96822}

\altaffiltext{4}{AIM, CNRS, Universit\'{e} Paris Diderot, B\^{a}t. 709, CEA-Saclay, 91191 Gif-sur-Yvette Cedex, France}

%\altaffiltext{4}{School of Physics, Monash University, Clayton, Victoria 2800, Australia}

%\altaffiltext{6}{Harvard Smithsonian Center for Astrophysics, 60 Garden Street, Cambridge MA 02138}

%\altaffiltext{7}{W. M. Keck Postdoctoral Fellow at the Harvard Smithsonian Center for Astrophysics}

%\altaffiltext{8}{Steward Observatory, Department of Astronomy, University of Arizona, 933 N. Cherry Avenue, Tucson AZ 85721}

%\altaffiltext{10}{Jet Propulsion Laboratory, California Institute of Technology, Mail Stop 169-527, Pasadena, CA 91109}

\begin{abstract}

Using data from the \textit{Spitzer Space Telescope}, we analyze the mid-infrared (3-70 $\micron$) spectral energy distributions of dry merger candidates in the \bootes \ field of the NOAO Deep Wide-Field Survey.  These candidates were selected by previous authors to be luminous, red, early-type galaxies with morphological evidence of recent tidal interactions.  We find that a significant fraction of these candidates exhibit 8 and 24 $\micron$ excesses compared to expectations for old stellar populations.  We estimate that a quarter of dry merger candidates have mid-infrared-derived star formation rates greater than $\sim$1 M$_{\odot}$ yr$^{-1}$.  This represents a ``frosting'' on top of a large old stellar population, and has been seen in previous studies of elliptical galaxies.  Further, the dry merger candidates include a higher fraction of starforming galaxies relative to a control sample without tidal features.  We therefore conclude that the star formation in these massive ellipticals is likely triggered by merger activity.  Our data suggest that the mergers responsible for the observed tidal features were not completely dry, and may be minor mergers involving a gas-rich dwarf galaxy.

 \end{abstract}


\section{Introduction}
\label{sec:intro}

In hierarchical models of galaxy formation, massive galaxies form via mergers of smaller galaxies.  Thus, the most massive galaxies, which tend to be ellipticals, should have formed most recently \citep[$z<1$; e.g.][]{White91,Cole00}.  In seeming contrast with this picture, studies of the stellar populations and scaling relations in samples of elliptical galaxies imply that the bulk of their stars formed significantly earlier, at $z \gsim 2$ \citep[e.g.][]{Djorgovski87,Bower92b,Bender93,Kuntschner00,Eisenhardt08}.  This discrepancy can be resolved if massive ellipticals form via \textit{dissipationless} mergers of red, bulge-dominated galaxies at low redshift.  These mergers would involve little to no gas, and are colloquially referred to as ``dry'' mergers.

Hierarchical scenarios in which a significant fraction of massive elliptical galaxies formed via dry mergers have been presented in a number of theoretical works \citep[e.g.][]{kauffmann00,khochfar03,khochfar09,deluciablaizot07}.  The observational evidence is mixed, with both supporting \citep[e.g.][]{vandokkum99,Bell04,Bell05,Bell06,vandokkum05,Tran05,Brown08} and contradictory \citep{Cimatti06, Scarlata07,Donovan07} studies.  

In this paper, we focus on the study of \citet{vandokkum05} (hereafter vD05), who analyzed the frequency of tidal distortions among an optically-selected sample of nearby ($z \approx 0.1$) bright red galaxies.  He found that 53\% of the entire color-selected sample shows morphological evidence of tidal interactions.  Further, this ratio rises to 71\% when considering only the bulge-dominated early-type galaxies in the sample.  vD05 concludes that the majority of today's most luminous field elliptical galaxies were assembled at low redshift through major dry mergers.

However, N-body simulations of binary galaxy mergers analyzed by \citet{Feldmann08} show that the morphologies of the tidal features seen in the vD05 sample cannot be reproduced by major dry mergers.  Instead, the observations are better explained by massive elliptical galaxies accreting much lower mass disk-dominated galaxies.  They also find that tidal features arising from disk accretion events last significantly longer than major elliptical-elliptical dry mergers (1-2 Gyr compared to a few hundred million years).  This pushes the primary epoch of elliptical mass assembly back to $z > 1$.  These simulations do not include gas.  However, if the mass of the accreted galaxy is small, and the lifetime of the tidal signatures is large, then \citet{Feldmann08} estimate that any stars formed during the interaction could have reddened with age sufficiently to match the observed colors. In a complementary study, \citet{Kawata06} use a cosmological N-body simulation including gas to show that a minor merger can result in a red elliptical galaxy displaying red tidal features, similar to some of the objects selected by vD05, if AGN heating is taken into account.

If the \citet{Feldmann08} and \citet{Kawata06} scenarios are correct, and the vD05 red galaxy tidal features are due to the accretion of a low mass, possibly disk-dominated galaxy, then the accreted companion could contain a significant reservoir of gas, which should be accompanied by dust.  \citet{Whitaker08} analyze $V_{606}$ and $I_{814}$ HST/ACS and WFPC2 images of 31 of the bulge-dominated red sequence galaxies presented by vD05 and found that only 10\% show evidence for dust based on their spatially-resolved colors.  Assuming a simple relation between dust and gas mass, they conclude that red mergers in the nearby Universe mostly involve early-type galaxies with little gas.

In contrast, \citet{Donovan07} examine 20 early-type galaxies known to be associated with neutral hydrogen.  Of these, 15 match the vD05 optical selection criteria.  The majority have $>$10$^8$~M$_{\odot}$ of HI.  In two cases, significant (up to 30$-$40~M$_{\odot}$~yr$^{-1}$) star formation is detected.  They conclude that red early-type galaxies are not the product of truly dry mergers.  \citet{Whitaker08} argue that this sample is not representative of massive ellipticals.  However, cold gas is observed in a large fraction of early-type galaxies in the nearby universe \citep{Morganti06,Combes07}.

The star formation activity within a subset of the vD05 sample was recently analyzed by \citet{SanchezBlazquez09} (hereafter SB09), who derived kinematics, stellar population absorption features, and ionization from emission lines.  They find that half of the sample with strong tidal features contain young stellar populations corresponding to 2\% of the baryonic mass of the galaxy, while a sample lacking interaction features does not contain detectable star formation.  They also find that the galaxies containing young stellar populations are supported by rotation, which is unexpected in remnants of major dry mergers \citep{Cox06, Naab03}.

Given these conflicting results regarding the gas content of optically selected dry mergers, we use data from the \textit{Spitzer Space Telescope} to revisit the question of whether the specific red mergers identified by vD05 are truly dry.  In \S\ref{sec:sample} we describe our multiwavelength data.  In \S\ref{sec:seds} we present the spectral energy distributions (SEDs) of a subsample of the vD05 galaxies and show that a significant fraction of the sources display mid-infrared excesses.  After arguing that the origin of these excesses is most likely star formation (\S\ref{sec:origin}), we estimate the implied star formation rates (SFRs) in \S\ref{sec:sfrs}, and discuss the contributions from AGN and AGB stars in \S\ref{sec:agn} and \S\ref{sec:agb}.  We then estimate the dust and gas content of these dry mergers in \S\ref{sec:gascontent}, discuss the results in \S\ref{sec:discussion},  and conclude in \S\ref{sec:conclusions}.  Throughout, we use ${\rm H}_0 = 70$~km~s$^{-1}$~Mpc$^{-1}$, $\Omega_{\rm m} = 0.3$, and $\Omega_\Lambda = 0.7$.  All quoted magnitudes are in the Vega system.

\section{The Sample and Survey Data}
\label{sec:sample}

In this paper, we investigate the properties of the sample of red galaxies presented by vD05.  Galaxies were selected by vD05 based on total $R$ magnitude and $B-R$ color.  The selection was tuned to yield nonstellar field objects with the colors and magnitudes of $L > L_{*}$ early-type galaxies at $0.05 < z < 0.2$.  The vD05 sample consists of 116 red galaxies selected from the optical imaging of the \bootes \ field of the NOAO Deep Wide-Field Survey \citep[NDWFS;][]{Jannuzi99}, and 10 galaxies selected from the Multiwavelength Survey by Yale-Chile \citep[MUSYC;][]{Gawiser06}.  The relative numbers reflect the relative areas of the parent surveys.  In this paper, we analyze the SEDs of the \bootes \ sample.

The \bootes \ field of the NDWFS has been observed with a variety of
telescopes, at wavelengths ranging from the X-ray to the radio.  The
multiwavelength data sets used in this paper are the following.

{\bf Ground-based Optical Imaging:} The 9.3 deg$^2$ \bootes \ field of the NDWFS has been imaged in the $B_W$, $R$, $I$, and $K$ bands down to 5$\sigma$ point-source depths of $\approx$27.1, 26.1, 25.4, and 19.0 Vega mags, respectively\footnote{See http://www.noao.edu/noao/noaodeep/ for more
  information regarding the depth and coverage of the NDWFS.}.  These are the data used by vD05 to select 116 red
galaxies.  The optical photometry plotted in this paper was determined
using the images of the third data release (DR3) of the NDWFS,
smoothed to achieve a uniform 1.35$\arcsec$ FWHM Moffat profile with a
$\beta$ parameter of 2.5. We used SExtractor AUTO magnitudes
\citep{Bertin96}.  

%{\bf Groundbased Near-IR imaging:} Additional imaging at the $J$ and
%$K_s$ bands was obtained for 4.7 deg$^2$ of the \bootes \ field
%through the FLAMEX survey \citep{Elston06}.

{\bf \textit{Spitzer} IRAC Imaging:} As part of the \textit{Spitzer} Deep
Wide-Field Survey \citep[SDWFS;][]{Ashby09}, 10 square degrees of the
\bootes \ field have been mapped with the Infrared Array Camera
\citep[IRAC;][]{Fazio04} on board the \textit{Spitzer Space
  Telescope}.  The 5$\sigma$, 4$\arcsec$ diameter, aperture-corrected SDWFS limits are 18.77, 18.83, 16.50, and 15.82 Vega~mag at 3.6, 4.5, 5.8, and 8.0~$\micron$, respectively.  We measured SExtractor AUTO magnitudes using the SDWFS version 3.2 zeropoints.  All 116 objects that were selected by vD05 from the optical \bootes \ data were also observed by IRAC.

{\bf \textit{Spitzer} MIPS Imaging:} Approximately 8.74 deg$^2$ of the \bootes \ field have been imaged with the Multiband Imaging Photometer for \textit{Spitzer} \citep[MIPS;][]{Rieke04}.  The 1$\sigma$ point-source depths of the MIPS survey are 0.051, 5, and 18 mJy at 24, 70, and 160~$\micron$, respectively.  The data were reduced by the MIPS GTO team.  Only five of the 116 sources selected by vD05 from the optical imaging of the \bootes \ field lack MIPS coverage: 6-2707, 2-3102, 3-953, 4-567, 2-3070.  All MIPS photometry quoted in this paper refers to the emission associated with the main galaxy.  In some cases, we see 24 $\micron$ emission in the near vicinity of the galaxy even though it is not centrally located.  This emission may be associated with the tidal arms and/or may be associated with the merger.  This emission is not accounted for in our current discussion, but may represent additional star formation associated with the merger event.  Examples include 2-368, 3-601, 4-1975, 11-1732, 13-3813, 16-584, 17-2819, 21-837, 25-3572, and 27-984.

{\bf \textit{Chandra} X-ray Imaging:} As part of the X\bootes \ survey, 9.3 deg$^2$ of the \bootes \ field has been imaged at a depth of 5~ks with ACIS-I on the \textit{Chandra X-ray Observatory} \citep{Murray05,Kenter05,Brand06}.  The limiting flux, corresponding to 4 or more X-ray counts, is $f_{(0.5-7{\rm keV})} = 8.1 \times 10^{-15}$ erg~cm$^{-2}$~s$^{-1}$.  The X-ray detections are discussed in \S\ref{sec:xray}.

\begin{figure}
\epsscale{1}
\plotone{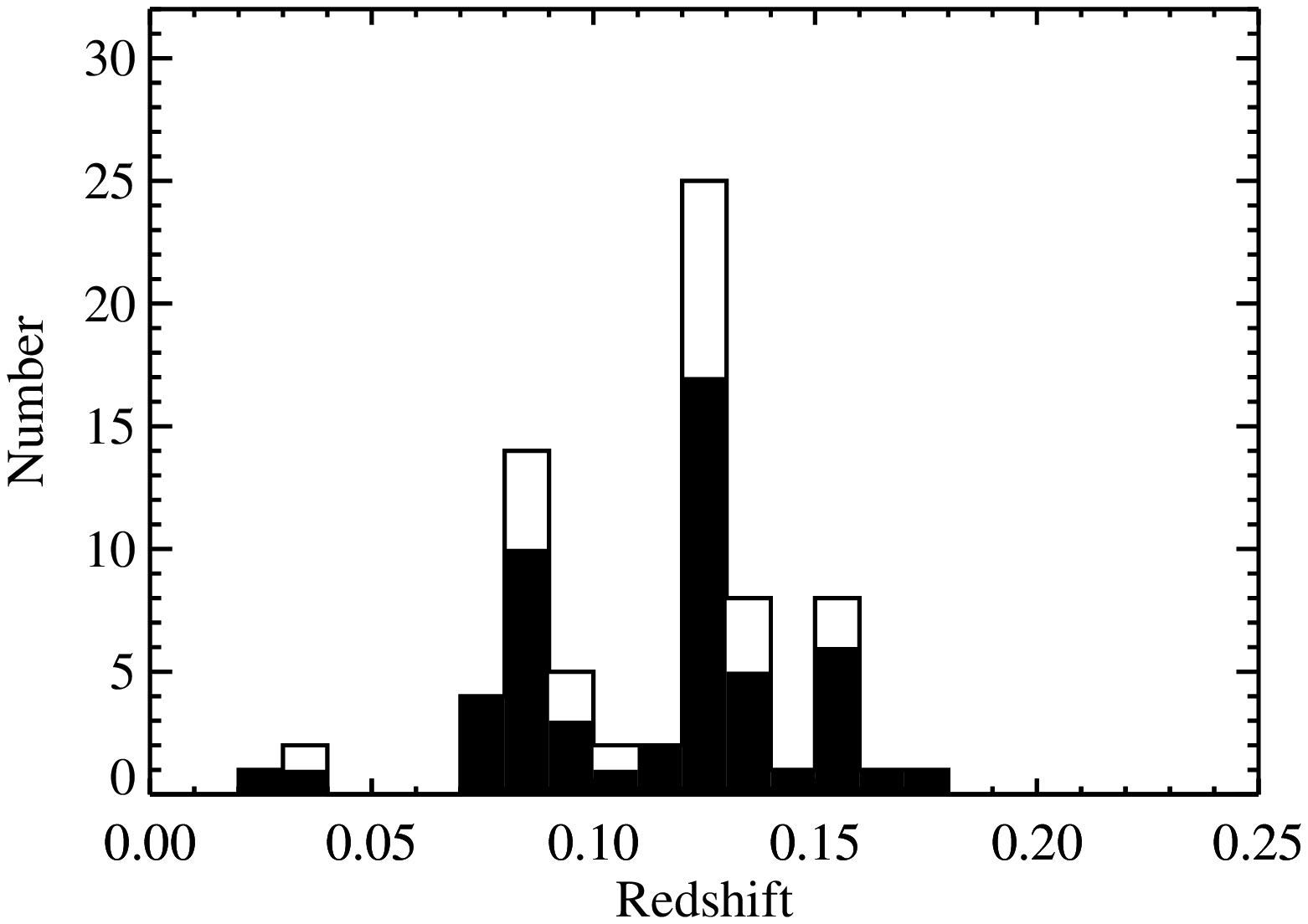}
\input{f1_caption}
\label{fig:zdist}
\end{figure}

{\bf \textit{SDSS} Optical Spectroscopy:} The \bootes \ field has also been observed as part of the Sloan Digital Sky Survey \citep[SDSS;][]{York00,Gunn06,Gunn98}.  We searched the SDSS DR7 \citep{SDSSDR7} database to find the optical spectra corresponding to our sources.  Of the 116 \bootes \ sources, we found optical spectra for 106.  Figure \ref{fig:zdist} shows their redshift distribution.  The sources lie in the range $0.08 < z < 0.17$, with a median redshift of $<z> = 0.1$.  

Many measurements and physical properties have been derived from the SDSS spectra in a homogenous way and made public by a team of SDSS researchers at the Max-Planck Institute for Astronomy at Heidelberg and Johns Hopkins University.  Although SDSS DR7 spectra are available, this value-added catalog is only complete through DR4 \citep{AdelmanMcCarthy06} at the time of writing.  Therefore, the reported optical SFRs, classifications, and stellar masses are measured from DR4 spectra\footnote{http://www.mpa-garching.mpg.de/SDSS/DR4}.

{\it\bf Samples:}  Our sample of red ellipticals is drawn from the set of galaxies identified as red ellipticals in the Bo\"otes field by vD05. We first selected all 75 galaxies  identified as red ellipticals (i.e., classified ``E/S0'', but not ``S0'' or ``S'') by vD05. In addition to the simple morphological classification, vD05 also visually classified galaxies into ``{\it tidal}'' classes, which take on four integer values (0 to 3) based on the following criteria: 0 for no tidal features; 1 for weak tidal features; 2 for strong tidal features; and 3 for evidence of an ongoing interaction with a neighboring galaxy. While the {\it tidal} class assigned to a galaxy is subjective and dependent on the depth of the imaging data available, it serves as a simple diagnostic to discriminate undisturbed ellipticals (i.e., those with {\it tidal}=0) from those which show some evidence of recent interaction (i.e., those with {\it tidal}$>$0). Of the sample of 75 ellipticals in the vD05 study, 21 are undisturbed ({\it tidal}=0) and 54 are dry merger candidates ({\it tidal}$>$0). Figure \ref{fig:zdist} shows the redshift distribution of the two samples. The redshift distributions are similar and suggest that the samples can be compared fairly. Table 1 lists the properties of the 54 dry merger candidates ({\it tidal}$>$0).

\section{Observed Mid-Infrared Excesses Among Dry Merger Candidates}
\label{sec:seds}

\begin{figure}
\epsscale{1}
\plotone{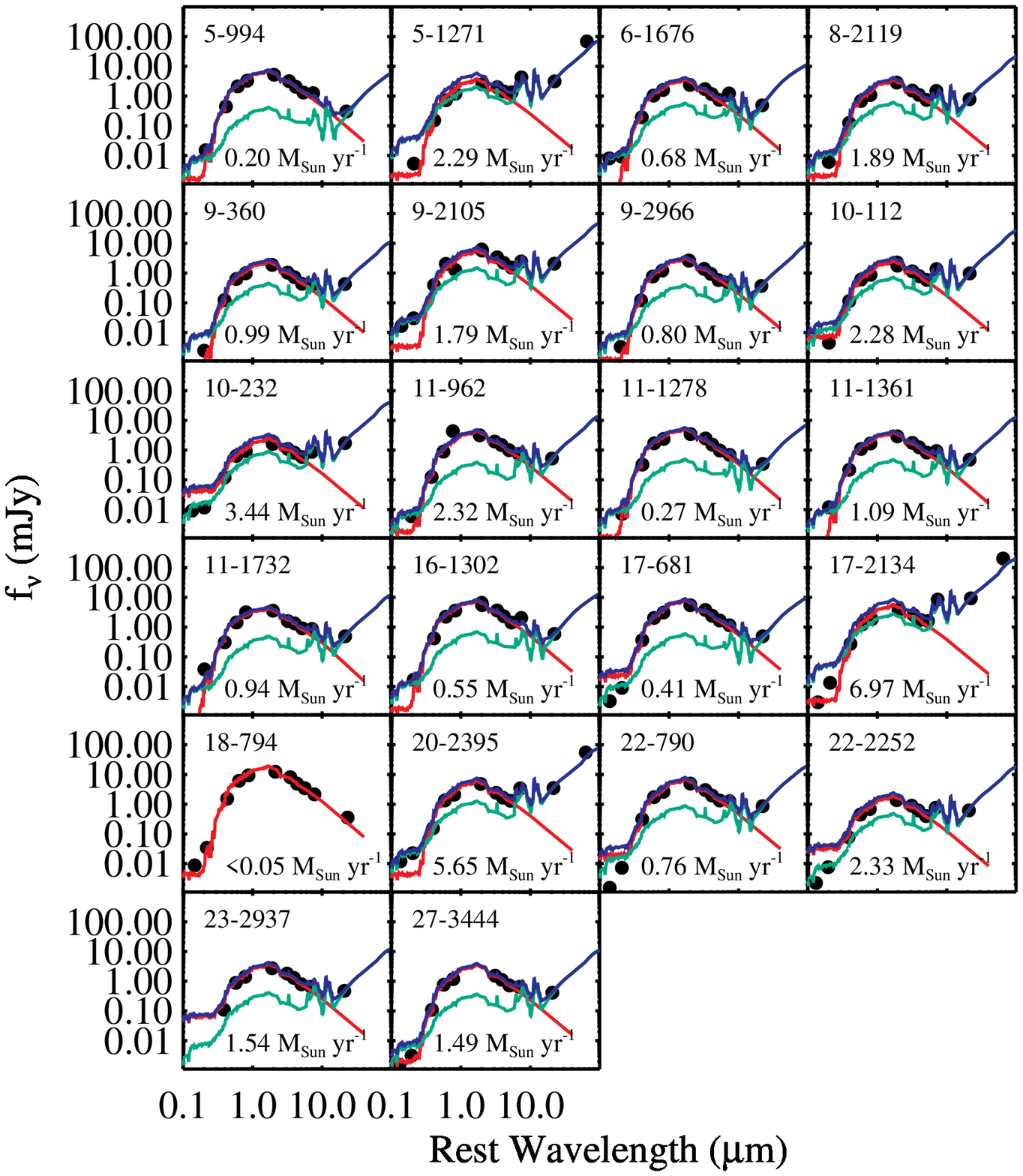}
\input{f2_caption}
\label{fig:seds}
\end{figure}

\begin{figure}
\epsscale{1}
\plotone{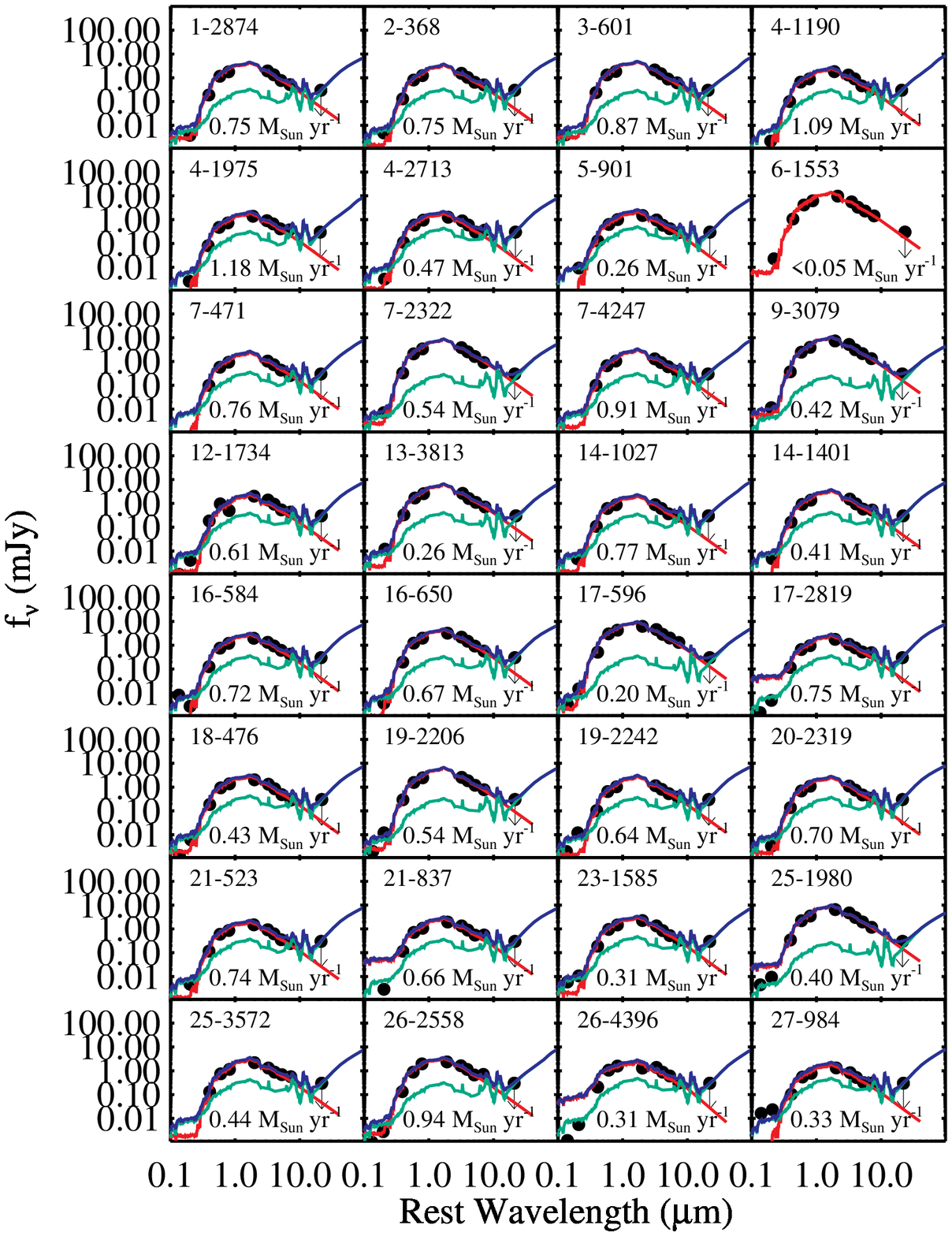}
\input{f3_caption}
\label{fig:moreseds}
\end{figure}

Using the multiwavelength survey data described in \S\ref{sec:sample}, we present the spectral energy distributions (SEDs) of the 54 \bootes \ dry merger candidates listed in Table \ref{table:sample}.  Figure \ref{fig:seds} shows the subset of 22 (41\%) that were detected at 24~$\micron$, while Figure \ref{fig:moreseds} shows the remaining 32 (59\%), which were undetected at 24~$\micron$.  Although a significant fraction of \bootes \ dry mergers were undetected in our 24~$\micron$ survey, all were detected at 8~$\micron$.  

Stellar population synthesis models computed using the isochrone synthesis code of \citet{Bruzual03} (hereafter BC03) were fit to the optical ($B_WRIK$) and near-infrared (IRAC 3.6, 4.5, 5.8~$\micron$) photometry for each object in Figures \ref{fig:seds} and \ref{fig:moreseds}. The models include three different
metallicities (0.008, 0.004, 0.02); have exponentially decreasing star formation rates with
$\tau=0.1,0.3,1,2,3,5,10,15,30$; and use a Chabrier IMF.  They do not include extinction. While the best-fitting BC03 models are good representations of the optical and near-infrared data, the photometry at longer wavelengths (IRAC 8~$\micron$, MIPS 24~$\micron$ and, in three cases, MIPS 70~$\micron$) varies significantly with respect to this model.  For example, Figures \ref{fig:seds} and \ref{fig:moreseds} show that the observed 24~$\micron$ flux density ranges from being consistent with an old stellar population (e.g.~18-794, 6-1553) to being over an order-of-magnitude in excess of it (e.g.~20-2395).  Large 8~$\micron$ excesses are also obvious in several sources (e.g.~5-1271, 10-112, 17-2134).

\begin{figure}
\epsscale{1}
\plotone{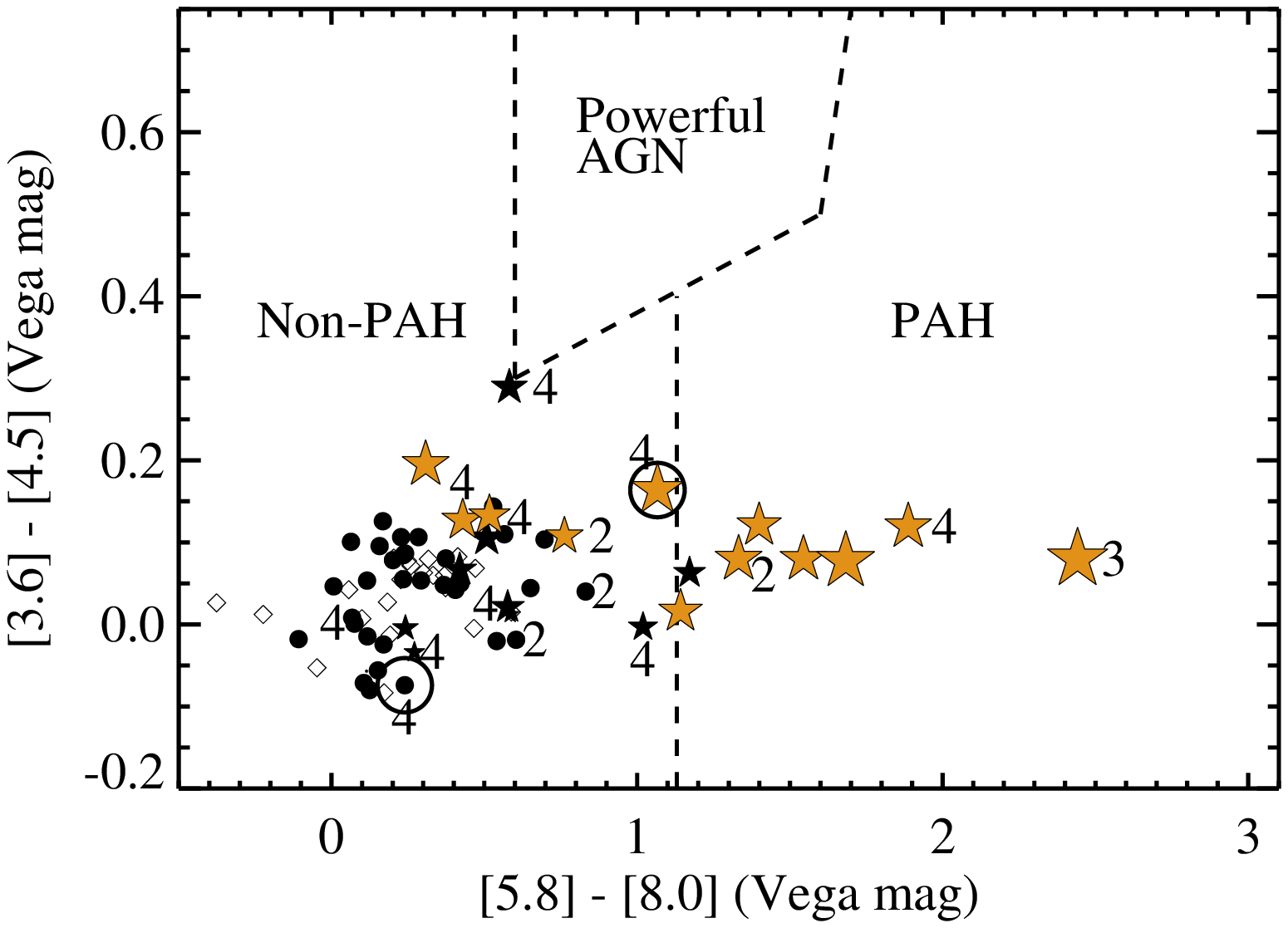}
\input{f4_caption}
\label{fig:iracirac}
\end{figure}

These excesses may be attributed to emission from some combination of 1) dusty star-forming regions; 2) dust heated by an active galactic nucleus (AGN); and 3) dust in the envelopes of stars on the Asymptotic Giant Branch (AGB).  The strong 8~$\micron$ excesses apparent in some sources in Figure \ref{fig:seds} imply polycyclic aromatic hydrocarbon (PAH) emission, which strongly suggests ongoing star formation.  In \S\ref{sec:origin}, we use this basic premise to argue that the observed mid-infrared excesses are dominated by dusty star formation.  However, we also consider the potential AGN and AGB contributions in \S\ref{sec:agn} and \S\ref{sec:agb}, respectively.  In \S\ref{sec:gascontent}, we go on to estimate the dust and gas content of the dry merger candidates under the simplifying assumption that all of the mid-infrared excesses are due entirely to star formation.  The SFRs used in Figures \ref{fig:seds} through \ref{fig:iracirac} all refer to values derived from the excess 24~$\micron$ emission, as described in \S\ref{sec:sfrs}.

\section{Origin of the Observed Mid-Infrared Excesses}
\label{sec:origin}

IRAC colors have been developed as a diagnostic to distinguish between star formation and AGN activity by several groups \citep{Lacy04,Sajina05,Stern05,Brand06}.  For star-forming galaxies at $z \approx 0.1$, the first three IRAC channels sample the Rayleigh-Jeans side of the blackbody contributed by old (cool) stars, while the fourth IRAC channel samples the rest-frame 7.7 $\micron$ PAH feature.  Thus, in star-forming galaxies, the [3.6]$-$[4.5] color is blue and the [5.8]$-$[8.0] color is red.  For passive galaxies with no ongoing star formation or AGN activity, there will be no PAH features, so both [3.6]$-$[4.5] and [5.8]$-$[8.0] are expected to be blue. For powerful AGN, both the stellar blackbody and the PAH features can be overwhelmed by emission from AGN-heated dust, leading to redder [3.6]$-$[4.5] and bluer [5.8]$-$[8.0] colors \citep{Brand09}.  

Figure \ref{fig:iracirac} shows an IRAC color-color diagnostic diagram with the \bootes \ dry merger candidates overplotted.  The AGN wedge was determined by \citet{Stern05} and was calibrated on optical spectroscopic diagnostics.  We adopt the \citet{Brand09} empirically-determined boundary between PAH (star-forming) and non-PAH (passive) emitting galaxies.  Based on this color-color diagram, many of the vD05 \bootes \ dry merger candidates exhibit colors consistent with PAH-emitting star-forming galaxies or passive galaxies with low levels of star formation activity.  Only one source (11-1732) lies near the AGN wedge in this diagram.  The mid-infrared colors do not rule out the presence of AGN in these galaxies (see \S\ref{sec:agn}), but they do imply that the observed mid-infrared excesses of the vast majority are dominated by star formation rather than AGN activity.

\section{Star Formation Rates}
\label{sec:sfrs}

\subsection{Infrared SFRs}
Figures \ref{fig:seds} and \ref{fig:moreseds} show the rest-frame SEDs of the \bootes \ dry merger candidates, as well as the BC03 model fits to the $B_WRIK$ and IRAC channels 1-3 photometry.  We estimate the infrared ($8-1000$ $\micron$) luminosity of each source from the \citet{Chary01} model that best fits the single data point represented by the 24 $\micron$ excess (the observed 24 $\micron$ flux density minus the expected 24~$\micron$ flux density of the BC03 model).  We find infrared luminosities in the range $1.2 \times 10^9 < {\rm
  L}_{8-1000 \micron} / {\rm L}_{\odot} < 4 \times 10^{10}$, corresponding to SFRs in the range $ 0.2 < {\rm SFR} / [{\rm M}_{\odot} {\rm yr}^{-1}] < 7$ \citep{Kennicutt98}.

For those sources without 24~$\micron$ detections, we calculate upper limits to the infrared SFR.  Given the typical redshift ($z \approx 0.1$) of the sources and the depth of the MIPS imaging, we are able to detect SFRs greater than approximately 1~M$_{\Sun}$~yr$^{-1}$.  Of the total sample of 54 dry merger candidates, 22 have 24~$\micron$ detections. Of these, 12 have SFR$\ge$1~M$_{\odot}$~yr$^{-1}$.  In addition, we can rule out star formation rates greater than 1~M$_{\odot}$ yr$^{-1}$ in all but two of the 32 dry merger candidates with only upper limits at 24~$\micron$.  Therefore, 12-14 out of 54 dry merger candidates (22-26\%) have infrared-derived SFR$\ge$1~M$_{\odot}$~yr$^{-1}$.

The infrared-derived SFRs and limits are listed in the second column of Table \ref{table:sample}.  We caution the reader that these star formation rates carry large (systematic) uncertainties, stemming from our lack of direct knowledge about the far-infrared SED, as well as scatter in the conversion between far-infrared luminosity and SFR.  We also remind the reader that in several cases we have excluded 24~$\micron$ emission arising outside the main body of the galaxy and are therefore missing some of the star formation in these galaxies.

\subsection{Spectroscopic SFRs}
\label{sec:spectra}

As described in \S\ref{sec:sample}, many of the \bootes \ dry merger candidates were spectroscopically observed as part of SDSS.  \citet{Brinchmann04} have estimated SFRs for galaxies based on SDSS DR4 spectra.  First, the spectra were classified according to the BPT diagram \citep[][\S\ref{sec:agn}]{Baldwin81}.  For galaxies classified as Star-Forming, \citet{Brinchmann04} modeled all of the emission lines to determine the SFR.  For Low SNR Star-Forming galaxies, the observed H$\alpha$ luminosity was used to calculate the SFR.  For the galaxies classified as AGN, Composite, or Unclassifiable, the measured values of D4000 were used to estimate SFRs.  All SFRs were aperture corrected.  The resulting SFRs are listed in Table \ref{table:derived}.  The stellar mass estimates (LGM in the \citet{Brinchmann04} catalog) are also presented in Table \ref{table:derived}.

SDSS-derived specific SFRs are available for 32 \bootes \ dry merger candidates classified as Star-Forming, Low SNR Star-Forming, AGN, Composite, or Unclassifiable. The values range from $6.2 \times 10^{-13}$ to $2.9 \times 10^{-11}$~yr$^{-1}$.  Of these 32, most (81\%) have specific SFRs less than $1 \times 10^{-11}$ yr$^{-1}$, making them typical of quiescent galaxies detected at 24 $\micron$ \citep{Salim09}.  However, 19\% (5-1271, 10-232, 11-1732, 17-2134, 22-2252, 27-3444) have specific SFRs exceeding $1~\times~10^{-11}$~yr$^{-1}$, making them transition objects between quiescent and star forming.  All of these have significant 24~$\micron$ excesses, but all have SDSS-derived stellar masses that are completely consistent with those of the quiescent galaxies. Of the six transition objects, Table \ref{table:sample} reveals that three have weak tidal features ($tidal = 1$), one has strong tidal features ($tidal = 2$), and two show evidence of an ongoing interaction with another galaxy ($tidal = 3$). Based on this small sample, we conclude that dry merger candidates with high spectroscopically-derived specific SFRs show diversity in the strength of their tidal features.

\begin{figure}
\epsscale{0.7}
\plotone{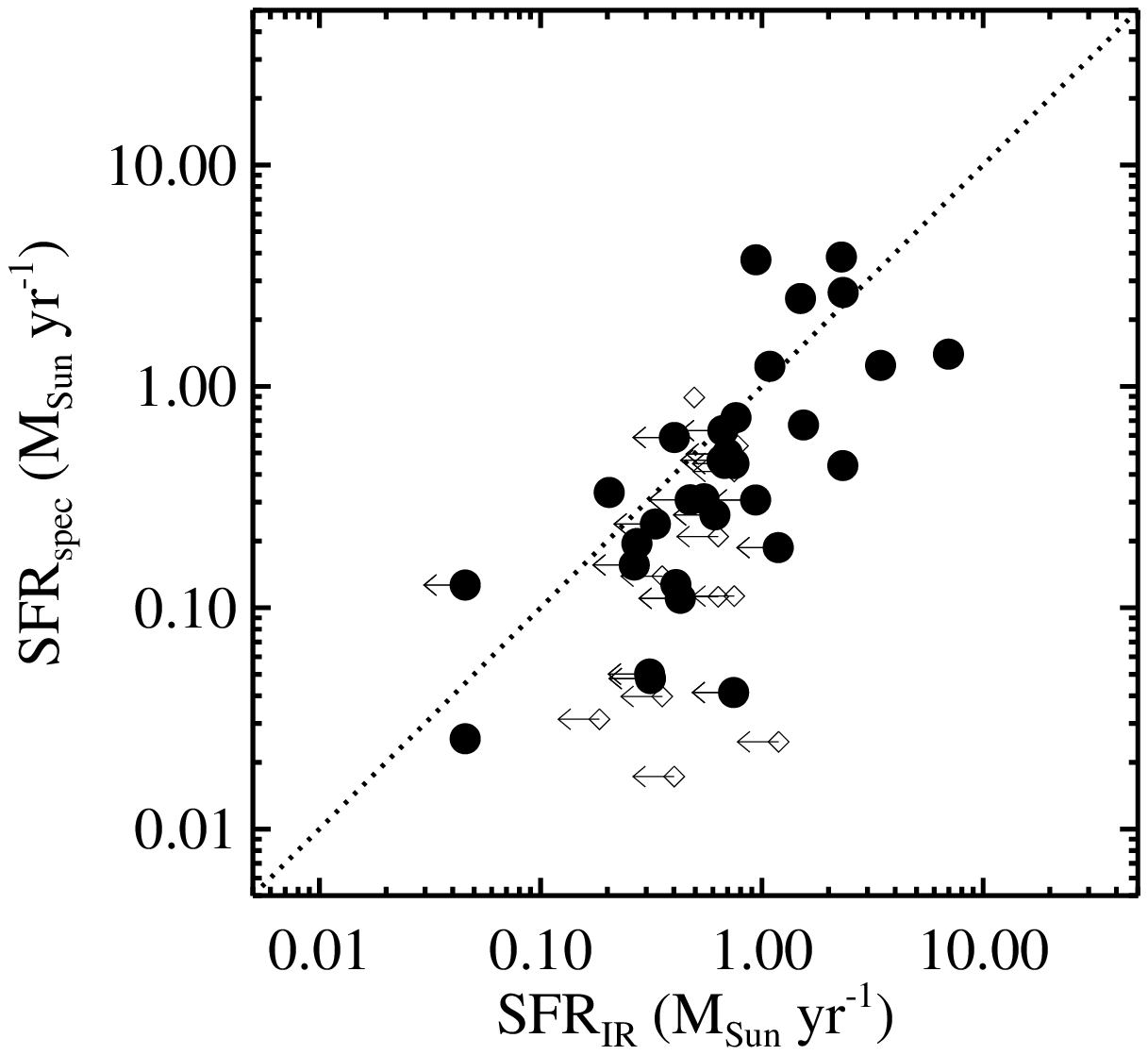}
\input{f5_caption}
\label{fig:irvspec}
\end{figure}

Figure \ref{fig:irvspec} shows the SDSS-derived SFRs versus the MIPS-derived SFRs.  The two are broadly correlated at a high level of significance according to the Spearman correlation coefficient ($\rho = 0.73$ when neglecting upper limits in the IR-derived SFRs).  On average, the MIPS-derived SFRs are a factor of $\approx$1.4 higher than the SDSS-derived SFRs, a discrepancy which may be due to dust-obscuration.

\section{AGN Content of Dry Merger Candidates}
\label{sec:agn}

Having identified star formation as the \textit{dominant} source of the mid-infrared excesses observed in Figure \ref{fig:seds}, we now investigate the AGN content of the \bootes \ dry mergers.  In particular, we consider  X-ray luminosity and optical spectroscopic line ratios.

\subsection{X-ray Luminosity}
\label{sec:xray}

The X-ray luminosity of a dry merger candidate can indicate its level of AGN activity.  The X\bootes \ survey has a limiting depth of $f_{(0.5-7{\rm keV})}~=~8.1~\times~10^{-15}~{\rm erg}~{\rm cm}^{-2}~{\rm s}^{-1}$.  At $z=0.1$, this corresponds to an X-ray luminosity of $2~\times~10^{41}~{\rm erg}~{\rm s}^{-1}$.  \citet{Grimm03} find that the X-ray luminosity of a galaxy is related to its SFR through the following relation:

\begin{equation} {\rm SFR}~[{\rm M}_{\odot}~ {\rm yr}^{-1}] = \frac{{\rm L}_{2-10 {\rm keV}}}{6.7 \times 10^{39}~{\rm erg~s}^{-1}} \end{equation}

\noindent for $L_{2-10{\rm keV}} \gsim 3 \times 10^{40}$~erg~s$^{-1}$.  This relation yields an X-ray derived SFR of $\approx$30 M$_{\odot}$~yr$^{-1}$ at the detection limit of the X\bootes \ survey.  This limiting SFR is a factor of $\approx$4 higher than the highest mid-infrared SFR derived for any of these sources (see \S\ref{sec:sfrs}), so any dry merger candidate detected in the X\bootes \ survey likely hosts an AGN.  

Using the X\bootes \ catalog of \citet{Brand06}, we found X-ray detections for only 5 of the red galaxies: 1-1403, 5-901, 5-2398, 10-232, and 26-5372.  Of these, only two (5-901 and 10-232) are dry merger candidates.  The object 5-901 was not detected at 24~$\micron$.  In contrast, 10-232 has $f_{\nu}(24 \micron)=1.71$~mJy, which translates to a fairly large SFR of $3.4$~$ {\rm M}_{\odot} {\rm yr}^{-1}$.  The X-ray luminosity calls into question whether the observed 24~$\micron$ excess for this source can be translated into a SFR, since AGN can also emit strongly at 24~$\micron$.  However, the SED shown in Figure \ref{fig:seds} appears to have an elevated 8~$\micron$ flux density, presumably from strong PAH features, indicating that star formation dominates the mid-infrared flux density.  The same conclusion can be drawn from Figure \ref{fig:iracirac}, where 10-232 (indicated by a gold star within a black circle) is very near the ``PAH'' region and well removed from the ``Powerful AGN'' region.

While we detect two X-ray luminous AGN among the dry merger candidates, the mid-infrared emission of these sources does not appear to be strongly affected by the AGN.

%, with derived star formation rates of 0.99 and 3.40
%M$_{\odot}$~yr$^{-1}$, respectively.

\subsection{BPT Diagram}
\label{sec:bpt}

The hard ionizing radiation of AGN results in optical emission line ratios distinct from those observed in star-forming regions.  For example, high [\ion{O}{3}]/H$\beta$ and [\ion{N}{2}]/H$\alpha$ ratios have been used to diagnose the presence of AGN in what is known as a BPT diagram \citep{Baldwin81}.  \citet{Brinchmann04} use 3$\arcsec$ diameter SDSS DR4 fiber spectroscopy to classify targeted galaxies into the following categories, which are numbered as in Table \ref{table:derived}.

{\bf Star-Forming (1):} The objects with ${\rm SNR} > 3$ in all four BPT
lines that have line ratios consistent with star formation.

{\bf Low SNR Star-Forming (2):} Galaxies with ${\rm SNR} > 2$ in H$\alpha$
that have not been classified as SF, AGN, or Composite.

{\bf Composite (3):} The objects with ${\rm SNR} > 3$ in all four BPT
lines for which up to 40\% of the H$\alpha$ luminosity has an AGN
origin.

{\bf AGN (4):} The objects with ${\rm SNR} > 3$ in all four BPT lines for
which a substantial AGN contribution is required to reproduce the BPT
line fluxes.  In addition, galaxies with AGN-like values of
[\ion{N}{2}]$\lambda$6584/H$\alpha$ and ${\rm SNR} > 3$ in the [\ion{N}{2}]$\lambda$6584 and H$\alpha$
lines but ${\rm SNR} < 3$ in either of the [\ion{O}{3}]$\lambda$5007 or H$\beta$
lines.

{\bf Unclassifiable (-1):} Those galaxies that cannot be classified using
the BPT diagram, typically because they have no or very weak emission
lines.

Of the 32 classified sources, four are classified as low SNR star-forming, 10 are classified as AGN, one is classified as Composite, and 17 are Unclassifiable.  Recall that these classifications are based on 3$\arcsec$ diameter fiber spectra.  Thus, the Unclassifiable sample likely includes galaxies which do not have star formation in the central bulge but may have star formation in the disk.  Similarly, while the optical emission lines indicate the presence of an AGN in some objects, this does not mean that the mid-infrared excess in these objects is \textit{dominated} by AGN activity.  The IRAC ratios presented in \S\ref{sec:sfrs} and Figure \ref{fig:iracirac} indicate that the objects with the largest mid-infrared excesses are dominated by star formation in the mid-infrared.  

Recently, SB09 spectroscopically classified 24 of the galaxies listed in our Tables \ref{table:sample} and \ref{table:derived}.  Since they had insufficient wavelength coverage to use the same line diagnostics as \citet{Brinchmann04}, they relied on [\ion{O}{2}]$\lambda$3727, H$\beta$, and [\ion{O}{3}]$\lambda$5007.  Their results are listed in column 5 of Table \ref{table:derived}.  There are 12 objects for which both SDSS and SB09 classifications exist.  Of these, six were listed by SB09 as ``?'', meaning there was insufficient wavelength coverage to measure enough lines for classification.  Of the remainder, three (12-1734, 17-681, 22-790) were consistently classified as having no or very weak emission lines; 10-232 was consistently classified as an AGN, and only two (6-1676 and 11-1732) had inconsistent diagnoses (-1 versus Seyfert and 4 versus LINER).

\section{AGB contribution to mid-infrared emission}
\label{sec:agb}

In Figures \ref{fig:seds} and \ref{fig:moreseds}, we model an old stellar population with BC03 fits.  However, such a stellar template may not accurately account for the mid-infrared emission from an old stellar population.  \citet{Kelson2010} recently showed that the large amount of near-infrared light expected from the thermally pulsating asymptotic giant branch (TP-AGB) phase \citep{Maraston05, Bruzual09, Conroy09}, when coupled with the observed mid-infrared fluxes of TP-AGB stars in our own galaxy, imply that a significant amount of mid-infrared flux could come from a $\sim$1 Gyr old stellar population.  This timescale is on the order of the time that a disk accretion event would have occurred according to \citet{Feldmann08} and thus, some mid-infrared emission could come from the stars that formed at or around the time of the accretion event itself. To test this effect, we have obtained updated models (commonly referred to as CB07 models in the literature) which use the prescription of \citet{Marigo07} and \citet{Marigo08} for the TP-AGB evolution of low- and intermediate-mass stars (S. Charlot 2011, private communication). We find that the resulting infrared-derived SFRs are very similar, and identical in many cases. We therefore conclude that for this population, the contribution of TP-AGB stars to the mid-infrared flux is insignificant.

\section{Dust and Gas Masses of Dry Merger Candidates}
\label{sec:gascontent}

The red optical colors of the \bootes \ dry merger candidates can be modeled by an old stellar population.  However, they could also be consistent with dust-extincted younger stars.  The dust mass of a galaxy can be inferred from its measured submillimeter flux.  Although submillimeter photometry is currently unavailable for these dry merger candidates, we can extrapolate based on the measured 24 $\micron$ flux densities.  Based on Figure \ref{fig:iracirac}, we assume that AGN do not contribute significantly to the far infrared fluxes of these objects.  Therefore, we use the best-fit \citet{Chary01} template from \S\ref{sec:agb} to estimate the 350 $\micron$ flux density of each source detected at 24 $\micron$ (or an upper limit in cases where 24 $\micron$ observations were available but there was no detection).  We estimate the dust mass following \citet{Hughes97}:

\begin{equation}
M_{\rm dust} = \frac{1}{1+z} \frac{F_{350} d_L^2}{\kappa_d B(\nu, T_d)},
\end{equation}

\noindent where $d_L$ is the luminosity distance; $\kappa_d$ is the rest-frequency mass absorption coefficient interpolated from \citet{Draine03}; and $B(\nu, T_d)$ is the value of the blackbody function at the rest frequency $\nu$ and a temperature $T_d$, which is taken to be 45 K.  Estimated dust masses are listed in Table \ref{table:derived}.  The dust masses for the dry merger candidates detected at 24 $\micron$ range from (0.3-10) $\times$ 10$^6$ M$_{\odot}$, with a mean of 3 $\times$ 10$^6$ M$_{\odot}$.  The dust mass upper limits for the dry merger candidates undetected at 24 $\micron$ range from (0.3-2) $\times$ 10$^6$ M$_{\odot}$, with a mean of 1.6 $\times$ 10$^6$ M$_{\odot}$.  For the canonical gas-to-dust ratio of 100, these dust masses correspond to gas masses ranging from (3-100) $\times$ 10$^7$ M$_{\odot}$ for the dry merger candidates detected at 24 $\micron$ and upper limits ranging from (3-20) $\times$ 10$^7$ M$_{\odot}$ for the dry merger candidates undetected at 24 $\micron$. Adopting a dust temperature of 30 K instead of 45 K would increase these estimates by a factor of two. 

Our \textit{Spitzer} observations reveal dust that went undetected in the HST analysis of \citet{Whitaker08}.  The 18 sources for which HST imaging is available are indicated in Table \ref{table:sample}.  Of these, we detected a 24 $\micron$ excess in half: 9-360, 9-2105, 11-962, 16-1302, 17-681, 17-2134, 22-790, 22-2252, 27-3444.  Note that 17-2134 has a particularly large 24 $\micron$ excess, corresponding to a SFR of 7~M$_{\odot}$~yr$^{-1}$ if all of it is attributed to star formation.  Interestingly, \citet{Whitaker08} find evidence for dust in 25-1980, which shows no mid-infrared excess.  Thus it seems that both the optical and mid-infrared imaging are necessary for a full census of the dust content of this population.

\begin{figure}
\epsscale{1}
\plotone{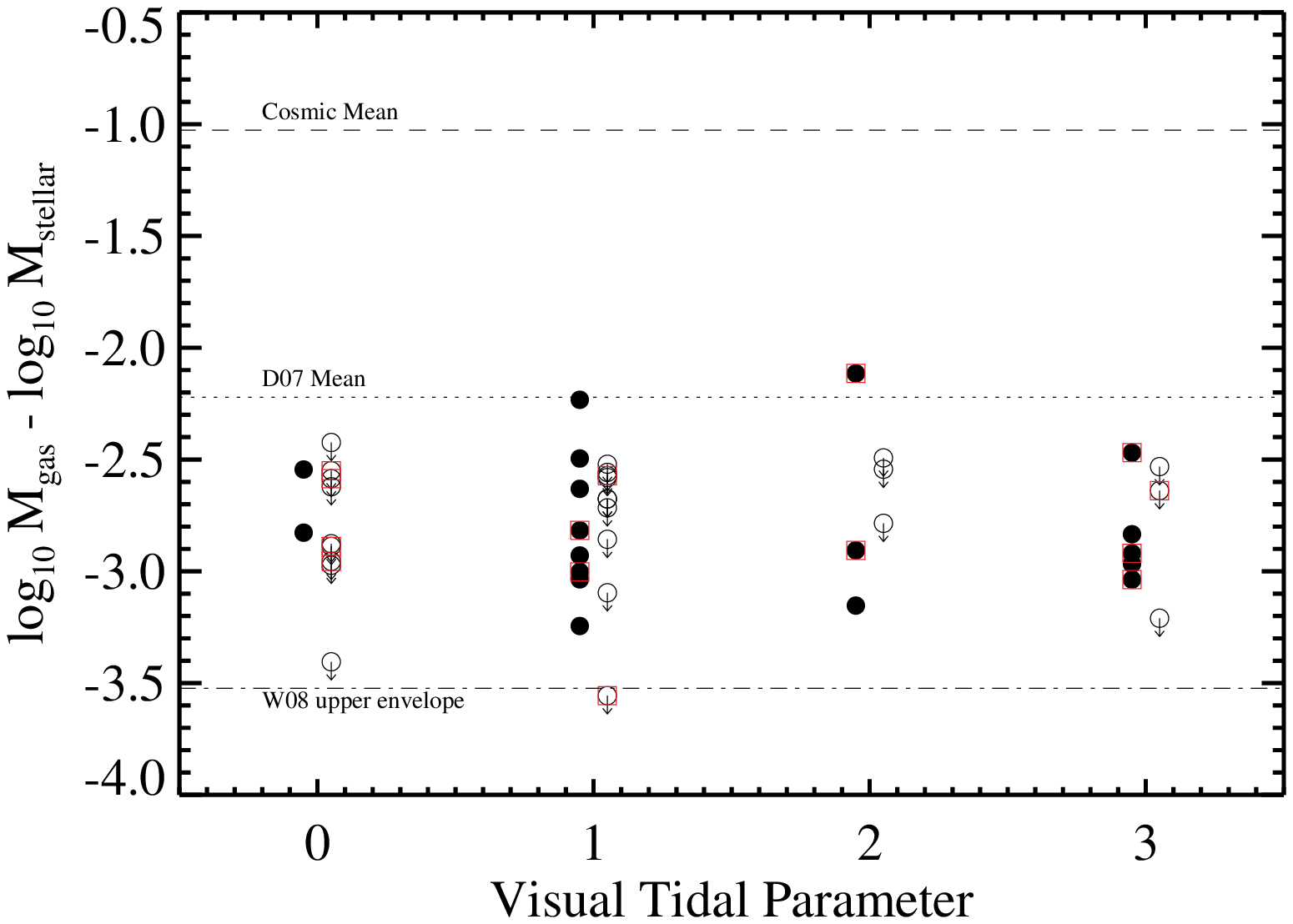}
\input{f6_caption}
\label{fig:whitaker}
\end{figure}

Figure \ref{fig:whitaker} shows the gas-to-stellar mass ratio versus vD05 \textit{tidal} parameter for the subset of the dry merger candidates for which we were able to calculate all quantities.  For comparison, we show the cosmic mean, the \citet{Donovan07} sample mean, and the \citet{Whitaker08} upper envelope, all taken from \citet{Whitaker08}.  The measured gas-to-stellar mass ratios derived by \textit{Spitzer} are intermediate between those derived for an overlapping sample by \citet{Whitaker08} and those derived for an analogous sample of early-type galaxies with HI emission by \citet{Donovan07}.  All are significantly below the cosmic mean.  Although our gas mass estimates are significantly higher than those calculated from optical images, we still find that gas makes up less than 1\% of the baryonic mass in these ellipticals.

A caveat to Figure \ref{fig:whitaker} are the uncertainties in calculating the gas masses.  These errors arise from uncertainties in the extrapolation to the far-infrared, the dust temperatures, the mass absorption coefficient, and the gas-to-dust ratio.  Considering only the last, we would have to adopt a gas-to-dust ratio of 10 to match the gas-to-stellar mass ratios of \citet{Whitaker08} and a gas-to-dust ratio of 250 to match to gas-to-stellar mass ratios of \citet{Donovan07}.

\section{Discussion}
\label{sec:discussion}

We have detected $>$1~M$_{\odot}~{\rm yr}^{-1}$ of star formation in $\approx$25\% of the massive dry merger candidates discovered by vD05 in \bootes.  Given the mass already existing in old stars, this represents a ``frosting'' of star formation, similar to what has been seen in previous studies of elliptical galaxies \citep[e.g.][]{Trager00, Gebhardt03}.  \citet{Kormendy09} have argued that the cuspy cores observed in the most luminous ellipticals may result from the scouring caused by binary black holes in dry mergers. The higher redshift of our sample and the lack of sufficiently high spatial resolution data preclude a determination of the nature of the central light profile. However, the absolute magnitude limit on the vD05 sample is approximately 1.5 mags fainter than the cuspy cores observed by Kormendy et al. Based on the SED combined with a number of assumptions, we estimate that these ellipticals are associated with on the order of 10$^8$ M$_{\odot}$ in gas.

\begin{figure}
\epsscale{0.7}
\plotone{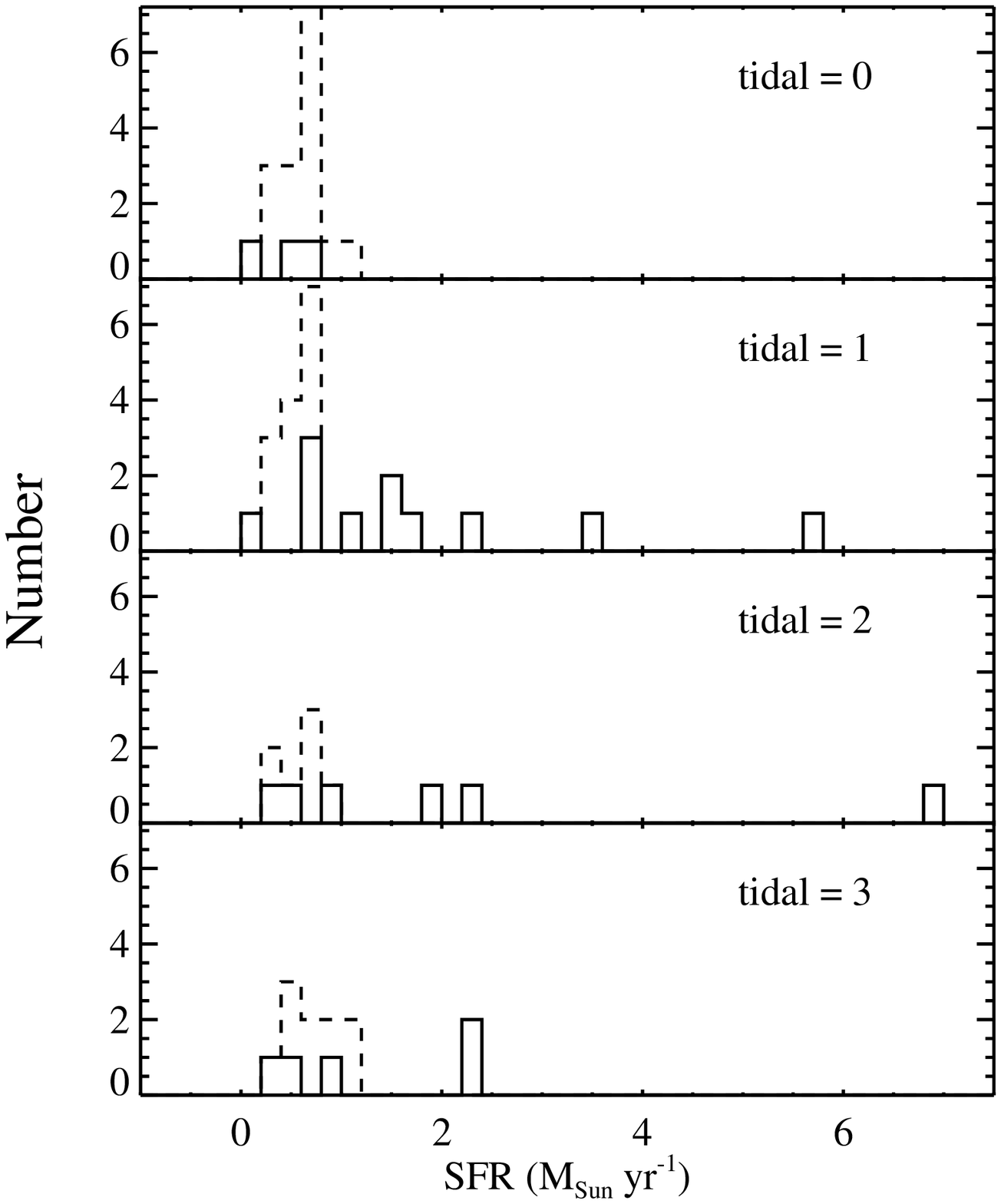}
\input{f7_caption}
\label{fig:morphplot}
\end{figure}

What is the origin of this gas?  We favor a scenario in which gas was delivered to the dry merger candidates via a merger.  The strongest evidence for this is that the residual star formation tends to be found in the sources with morphological evidence of a recent merger.   Figure \ref{fig:morphplot} shows the distribution of infrared-derived SFRs in bins of the $tidal$ parameter tabulated by vD05.  Distributions are presented both for the \bootes \ dry merger candidates and for a control sample consisting of the subset of the red galaxy sample with early type morphologies and no tidal features.  Galaxies forming stars at a rate greater than about 1 M$_{\odot}$ yr$^{-1}$ tend have a $tidal$ parameter greater than 0.  The observation that the mid-infrared emission is related to the $tidal$ designation supports the hypothesis that the star formation is due to interaction-driven activity. In a GALEX study of these same sources, \citet{Kaviraj2010} also found that the dry merger candidates with the bluest GALEX colors tended to show signs of tidal interaction. Additionally, \citet{Donovan07} looked at HI data for galaxies that would fall into the dry merger sample, and found large reservoirs of gas.  They find that 12/15 red rogues exhibit signs of tidal interaction.  This is comparable to the fraction of red early-type galaxies found by vD05 to exhibit tidal features.

An alternative explanation is that gas expelled from AGB stars is already present in the dry merger candidates, and star formation is triggered by a merger, not fueled by one.  Can AGB stars expel enough gas to fuel star formation?  For a $10^{11}$~M$_{\odot}$ old stellar population, 0.15~M$_{\odot}$~yr$^{-1}$ is expected to be ejected by AGB stars \citep{Mathews03}.  Over a billion years, such a galaxy could accumulate $~$10$^8$~M$_{\odot}$ of gas, which is on the same order as what is observed.  If this were the case, then we would expect that even the galaxies without tidal features should have evidence for gas, even if it has not been triggered to form stars.  

\section{Conclusions}
\label{sec:conclusions}

We analyze the multiwavelength data available for vD05 dry merger
candidates in the NDWFS \bootes \ field.  We find:

\begin{itemize}

\item A significant fraction of the sources display mid-infrared
  (24~$\micron$) excesses over that expected from an old stellar
  population with the observed red optical colors.

\item Based on the mid-infrared IRAC colors indicating the presence of PAH emission, this infrared excess is likely due to emission from dust heated
  by star forming regions, rather than AGN-heated dust or AGB stars.

\item If the observed mid-infrared excesses are due to star formation,
  we estimate that a quarter of the \bootes \ dry merger candidates are
  forming stars at rates greater than 1 M$_{\odot}$ yr$^{-1}$.  This represents a ``frosting'' of star formation on top of a well-developed old stellar population.

\item Red early-type galaxies exhibiting tidal features are more likely to have star formation detectable in the mid-infrared than a control sample lacking tidal features.  This implies that a frosting of star formation in elliptical galaxies may be triggered by tidal interactions.

%\item SDSS-derived specific SFRs are available for about 60\% of the \bootes \ dry mergers.  Of these, 19\% have specific SFRs greater than $1 \times 10^{-11}$ yr$^{-1}$, which is larger than expected for quiescent galaxies.

\item We estimate gas masses in the range (3-100) $\times$ 10$^7$ M$_{\odot}$ for the dry merger candidates detected at 24~$\micron$ and upper limits ranging from (3-20) $\times$ 10$^7$ M$_{\odot}$ for the dry merger candidates undetected at 24 $\micron$.  

\item Based on the observed 24~$\micron$ emission, and assuming the \citet{Chary01} star-forming templates, we predict the 70 $\micron$ flux densities of the dry merger candidates.  The predicted 70 $\micron$ flux densities are shown in Table \ref{table:sample}.  Only three sources have a \textit{Spitzer} 70 $\micron$ detection.  For these three sources, the observed 70 $\micron$ flux density is a factor of 1.5-2 times greater than the template prediction.  Nevertheless, these template predictions provide useful guidelines for future Herschel observations.

\end{itemize}

\acknowledgments
\section{Acknowledgments}
\label{sec:acknowledgments}

AD thanks the SSC/Caltech for its gracious hospitality during summer 2009, when much of this paper was written.  EC was supported through the Caltech Summer Undergraduate Research Fellowship program and the \textit{Spitzer} Enchanced Science Fund.  ELF is supported by the \textit{Spitzer} Fellowship Program through a contract with JPL/Caltech/NASA.  We thank John Moustakas for advice on using the SDSS spectra, and St\'{e}phane Charlot for providing the CB07 models. We also warmly thank Lee Armus, George Helou, Bradford Holden, Daniel Kelson, Francine Marleau, Samir Salim, and Nick Scoville for stimulating discussions pertaining to this work. Finally, we are grateful to the anonymous referee for providing useful feedback that improved this work.

This research is partially supported by the National Optical Astronomy Observatory which is operated by the Association of Universities for Research in Astronomy, Inc. (AURA) under a cooperative agreement with the National Science Foundation.

This work is in part based on observations made with the \textit{Spitzer Space Telescope}, which is operated by the Jet Propulsion Laboratory, California Institute of Technology under a contract with NASA.  Support for this work was provided by NASA through an award issued by JPL/Caltech.

The \textit{Spitzer}/MIPS survey of the \bootes \ region was obtained
using GTO time provided by the \textit{Spitzer} Infrared Spectrograph
Team (James Houck, P.I.) and by M. Rieke.

Funding for the SDSS and SDSS-II has been provided by the Alfred
P. Sloan Foundation, the Participating Institutions, the National
Science Foundation, the U.S. Department of Energy, the National
Aeronautics and Space Administration, the Japanese Monbukagakusho, the
Max Planck Society, and the Higher Education Funding Council for
England. The SDSS Web Site is http://www.sdss.org/.

The SDSS is managed by the Astrophysical Research Consortium for the
Participating Institutions. The Participating Institutions are the
American Museum of Natural History, Astrophysical Institute Potsdam,
University of Basel, University of Cambridge, Case Western Reserve
University, University of Chicago, Drexel University, Fermilab, the
Institute for Advanced Study, the Japan Participation Group, Johns
Hopkins University, the Joint Institute for Nuclear Astrophysics, the
Kavli Institute for Particle Astrophysics and Cosmology, the Korean
Scientist Group, the Chinese Academy of Sciences (LAMOST), Los Alamos
National Laboratory, the Max-Planck-Institute for Astronomy (MPIA),
the Max-Planck-Institute for Astrophysics (MPA), New Mexico State
University, Ohio State University, University of Pittsburgh,
University of Portsmouth, Princeton University, the United States
Naval Observatory, and the University of Washington.

{\it Facilities:} \facility{Spitzer(MIPS,IRAC,IRS)}
\facility{KPNO:2.1m(ONIS,SQUID,FLAMINGOS,FLAMINGOS-1)}
\facility{Mayall(Mosaic-1)}

\bibliographystyle{apj}
\bibliography{references}

\renewcommand{\baselinestretch}{1.9}

%%%FIGURE 1

%%%FIGURE 2

%%%FIGURE 3

%%%FIGURE 4

%%%FIGURE 5

%%%FIGURE 6

%%%FIGURE 7

% TABLE 1
\clearpage
\LongTables
\begin{landscape}
\tabletypesize{\tiny}
\begin{deluxetable}{clcccccccccccccc}
\tablecaption{Dry Merger Candidates}
\tablewidth{0pt}
\startdata \hline\hline
Object     & RA      &  Dec                            & tidal & z\tablenotemark{a}   & [3.6]      & [4.5]      & [5.8]      & [8.0]      & $f_{\nu}(24 \micron)$ & $f_{\nu, {\rm obs}}(70 \micron)$ & $f_{\nu, {\rm pred}}(70 \micron)$ & remark \\
           & (h:m:s) & ($\deg$:$\arcmin$:$\arcsec$)    &       &                      & (Vega mag) & (Vega mag) & (Vega mag) & (Vega mag) & (mJy)                & (mJy)                     & (mJy)                     & \\\hline

    1-2874 &      14:27:19.5 &      32:44:29.1 & 3 & 0.1333 & 12.90 & 12.82 & 12.93 & 12.73 & $<$0.3 & $<$25 & $<$3 & SB09 \\ 
     2-368 &      14:26:15.6 &      32:54:27.4 & 2 & 0.1315 & 13.15 & 13.10 & 13.13 & 12.76 & $<$0.3 & $<$25 & $<$3 & \nodata \\ 
     3-601 &      14:27:03.5 &      33:33:51.8 & 2 & 0.1514 & 12.79 & 12.71 & 12.79 & 12.55 & $<$0.3 & $<$25 & $<$2 & W08; SB09 \\ 
    4-1190 &      14:25:43.7 &      34:17:02.8 & 3 & 0.1563 & 13.33 & 13.24 & 13.50 & 13.26 & $<$0.3 & $<$25 & $<$3 & W08 \\ 
    4-1975 &      14:26:43.2 &      34:28:53.3 & 3 & 0.1602 & 13.73 & 13.58 & 13.62 & 13.09 & $<$0.3 & $<$25 & $<$3 & W08 \\ 
 &  &  &  &  &  &  &  &  &  &  &  &  &  \\ 
    4-2713 &      14:25:03.5 &      34:37:49.0 & 3 & 0.1009 & 13.74 & 13.70 & 13.95 & 13.12 & $<$0.3 & $<$25 & $<$3 & \nodata \\ 
     5-901 &      14:26:35.8 &      34:49:48.6 & 1 & 0.0767 & 13.65 & 13.73 & 13.61 & 13.37 & $<$0.3 & $<$25 & $<$3 & \nodata \\ 
     5-994 &      14:27:03.2 &      34:51:35.0 & 2 & 0.0775 & 12.37 & 12.36 & 12.36 & 11.78 & 0.30 & $<$25 & 2 & SB09 \\ 
    5-1271 &      14:25:41.1 &      34:54:47.4 & 1 & 0.0766 & 12.88 & 12.76 & 12.34 & 10.45 & 3.10 & 68 & 32 & \nodata \\ 
    6-1553 &      14:26:14.7 &      35:31:05.5 & 1 & 0.0284 & 11.75 & 11.83 & 11.75 & 11.62 & $<$0.3 & $<$25 & $<$4 & SB09 \\ 
 &  &  &  &  &  &  &  &  &  &  &  &  &  \\ 
    6-1676 &      14:25:55.1 &      35:32:17.6 & 1 & 0.0995 & 13.04 & 12.98 & 12.98 & 11.80 & 0.48 & $<$25 & 5 & SB09 \\ 
     7-471 &      14:29:06.0 &      32:21:32.9 & 1 & 0.1313 & 13.58 & 13.48 & 13.56 & 13.49 & $<$0.3 & $<$25 & $<$3 & \nodata \\ 
    7-2322 &      14:27:53.9 &      32:37:14.7 & 1 & 0.1299 & 12.16 & 12.12 & 12.13 & 12.12 & $<$0.3 & $<$25 & $<$2 & SB09 \\ 
    7-4247 &      14:28:12.8 &      32:54:30.5 & 3 & 0.1496 & 13.34 & 13.25 & 13.20 & 13.04 & $<$0.3 & $<$25 & $<$2 & \nodata \\ 
    8-2119 &      14:28:54.7 &      33:12:12.2 & 2 & 0.1330 & 13.16 & 13.04 & 12.99 & 11.59 & 0.77 & $<$25 & 8 & SB09 \\ 
 &  &  &  &  &  &  &  &  &  &  &  &  &  \\ 
     9-360 &      14:28:29.6 &      33:29:53.8 & 2 & 0.1286 & 13.59 & 13.48 & 13.56 & 13.05 & 0.43 & $<$25 & 4 & W08 \\ 
    9-2105 &      14:29:19.2 &      33:48:58.0 & 1 & 0.0840 & 12.30 & 12.28 & 12.19 & 11.05 & 2.06 & $<$25 & 20 & W08; SB09 \\ 
    9-2966 &      14:28:24.5 &      33:57:38.4 & 1 & 0.1280 & 13.27 & 13.20 & 13.25 & 12.83 & 0.37 & $<$25 & 3 & \nodata \\ 
    9-3079 &      14:29:14.3 &      33:59:22.4 & 2 & 0.1293 & 11.83 & 11.90 & 11.82 & 11.72 & $<$0.3 & $<$25 & $<$1 & SB09 \\ 
    10-112 &      14:29:12.9 &      34:04:41.4 & 2 & 0.1280 & 13.48 & 13.40 & 13.23 & 11.69 & 0.97 & $<$25 & 10 & SB09 \\ 
 &  &  &  &  &  &  &  &  &  &  &  &  &  \\ 
    10-232 &      14:30:03.2 &      34:06:51.3 & 1 & 0.1261 & 13.53 & 13.37 & 13.24 & 12.18 & 1.71 & $<$25 & 19 & SB09 \\ 
    11-962 &      14:29:09.7 &      34:52:13.2 & 3 & 0.1735 & 12.90 & 12.71 & 12.66 & 12.35 & 0.52 & $<$25 & 4 & double nucleus; W08 \\ 
   11-1278 &      14:29:30.2 &      34:57:23.7 & 3 & 0.0776 & 12.64 & 12.68 & 12.69 & 12.42 & 0.34 & $<$25 & 3 & SB09 \\ 
   11-1361 &      14:27:44.1 &      34:58:34.4 & 1 & 0.1287 & 13.01 & 12.91 & 12.86 & 12.10 & 0.48 & $<$25 & 5 & \nodata \\ 
   11-1732 &      14:28:55.5 &      35:04:56.7 & 3 & 0.1210 & 13.12 & 12.83 & 12.76 & 12.17 & 0.49 & $<$25 & 4 & SB09 \\ 
 &  &  &  &  &  &  &  &  &  &  &  &  &  \\ 
   12-1734 &      14:29:28.2 &      35:37:53.5 & 1 & 0.1164 & 13.27 & 13.23 & 13.34 & 12.93 & $<$0.3 & $<$25 & $<$3 & SB09 \\ 
   13-3813 &      14:32:29.1 &      32:46:20.0 & 2 & 0.0850 & 12.57 & 12.63 & 12.55 & 12.40 & $<$0.3 & $<$25 & $<$3 & SB09 \\ 
   14-1027 &      14:30:27.7 &      33:02:29.9 & 1 & 0.1336 & 13.76 & 13.68 & 13.75 & 13.38 & $<$0.3 & $<$25 & $<$3 & \nodata \\ 
   14-1401 &      14:31:34.8 &      33:06:18.5 & 3 & 0.0984 & 13.17 & 13.19 & 13.26 & 12.72 & $<$0.3 & $<$25 & $<$3 & W08; SB09 \\ 
    16-584 &      14:32:51.4 &      34:07:26.6 & 1 & 0.1269 & 13.32 & 13.21 & 13.19 & 12.96 & $<$0.3 & $<$25 & $<$4 & \nodata \\ 
 &  &  &  &  &  &  &  &  &  &  &  &  &  \\ 
    16-650 &      14:31:30.8 &      34:08:09.1 & 1 & 0.1282 & 12.78 & 12.72 & 12.74 & 12.63 & $<$0.3 & $<$25 & $<$3 & SB09 \\ 
   16-1302 &      14:30:55.7 &      34:14:28.2 & 2 & 0.0843 & 12.22 & 12.22 & 12.27 & 11.25 & 0.59 & $<$25 & 6 & W08; SB09 \\ 
    17-596 &      14:31:54.0 &      34:42:10.7 & 3 & 0.0840 & 11.97 & 12.00 & 11.90 & 11.73 & $<$0.3 & $<$25 & $<$1 & pair with 17-681; W08; SB09 \\ 
    17-681 &      14:31:55.3 &      34:42:40.6 & 3 & 0.0831 & 12.18 & 12.18 & 12.10 & 11.86 & 0.49 & $<$25 & 5 & pair with 17-596; W08; SB09 \\ 
   17-2134 &      14:30:48.3 &      34:56:06.0 & 2 & 0.0843 & 12.60 & 12.52 & 12.14 & 9.70 & 9.32 & 203 & 100 & W08; SB09 \\ 
 &  &  &  &  &  &  &  &  &  &  &  &  &  \\ 
   17-2819 &      14:32:16.8 &      35:02:48.9 & 1 & 0.1282 & 13.52 & 13.41 & 13.37 & 12.80 & $<$0.3 & $<$25 & $<$4 & \nodata \\ 
    18-476 &      14:32:03.5 &      35:16:23.5 & 1 & 0.0986 & 13.32 & 13.31 & 13.45 & 13.38 & $<$0.3 & $<$25 & $<$3 & \nodata \\ 
    18-794 &      14:31:12.8 &      35:19:55.1 & 1 & 0.0327 & 11.35 & 11.41 & 11.31 & 11.19 & 0.35 & $<$25 & 3 & \nodata \\ 
   19-2206 &      14:35:41.8 &      33:08:20.1 & 3 & 0.1210 & 12.60 & 12.47 & 12.46 & 12.30 & $<$0.3 & $<$25 & $<$3 & pair with 19-2242; W08; SB09 \\ 
   19-2242 &      14:35:42.4 &      33:08:22.4 & 3 & 0.1205 & 13.48 & 13.37 & 13.35 & 13.06 & $<$0.3 & $<$25 & $<$3 & pair with 19-2206; W08; SB09 \\ 
 &  &  &  &  &  &  &  &  &  &  &  &  &  \\ 
   20-2319 &      14:34:29.7 &      33:47:25.6 & 1 & 0.1238 & 13.51 & 13.46 & 13.48 & 13.19 & $<$0.3 & $<$25 & $<$3 & \nodata \\ 
   20-2395 &      14:33:37.5 &      33:48:13.8 & 1 & 0.1190 & 12.67 & 12.59 & 12.37 & 10.69 & 3.49 & 54 & 37 & \nodata \\ 
    21-523 &      14:33:56.8 &      34:09:00.7 & 2 & 0.1257 & 13.71 & 13.65 & 13.79 & 13.55 & $<$0.3 & $<$25 & $<$4 & \nodata \\ 
    21-837 &      14:33:51.6 &      34:13:01.2 & 2 & 0.1225 & 13.30 & 13.26 & 13.30 & 12.65 & $<$0.3 & $<$25 & $<$3 & \nodata \\ 
    22-790 &      14:33:27.1 &      34:43:42.1 & 1 & 0.0846 & 12.44 & 12.42 & 12.35 & 11.77 & 0.86 & $<$25 & 9 & W08; SB09 \\ 
 &  &  &  &  &  &  &  &  &  &  &  &  &  \\ 
   22-2252 &      14:33:37.5 &      34:58:23.2 & 3 & 0.1561 & 13.75 & 13.67 & 13.66 & 12.32 & 0.62 & $<$25 & 7 & W08 \\ 
   23-1585 &      14:33:12.2 &      35:30:04.8 & 1 & 0.0839 & 13.30 & 13.32 & 13.35 & 13.23 & $<$0.3 & $<$25 & $<$3 & W08 \\ 
   23-2937 &      14:34:49.1 &      35:42:47.3 & 1 & 0.1520 & 12.98 & 12.86 & 12.91 & 12.48 & 0.47 & $<$25 & 4 & \nodata \\ 
   25-1980 &      14:37:07.1 &      34:18:50.7 & 1 & 0.1220 & 11.94 & 11.94 & 11.92 & 11.85 & $<$0.3 & $<$25 & $<$2 & W08 \\ 
   25-3572 &      14:37:21.9 &      34:34:56.0 & 1 & \nodata & 13.33 & 13.28 & 13.06 & 12.64 & $<$0.3 & $<$25 & $<$3 & \nodata \\ 
 &  &  &  &  &  &  &  &  &  &  &  &  &  \\ 
   26-2558 &      14:36:42.0 &      34:55:14.5 & 3 & 0.1509 & 13.10 & 12.99 & 13.09 & 12.40 & $<$0.3 & $<$25 & $<$2 & \nodata \\ 
   26-4396 &      14:38:08.4 &      35:10:00.2 & 1 & 0.0830 & 13.27 & 13.28 & 13.20 & 13.31 & $<$0.3 & $<$25 & $<$3 & \nodata \\ 
    27-984 &      14:36:10.2 &      35:21:06.6 & 2 & 0.0852 & 13.82 & 13.84 & 13.89 & 13.29 & $<$0.3 & $<$25 & $<$3 & \nodata \\ 
   27-3444 &      14:37:20.3 &      35:38:16.7 & 1 & 0.1593 & 13.16 & 13.03 & 13.02 & 12.50 & 0.41 & $<$25 & 3 & W08 \\ 

\enddata 
\tablecomments{Subset of red galaxies from vd05 which are early type and have tidal features.  Redshifts are from SDSS DR7 and/or SB09.  The IRAC magnitudes are SExtractor AUTO magnitudes.  The MIPS fluxes are PSF-fitted total fluxes.  All limits are 5$\sigma$.  If W08 appears in the remarks, the source was included in the HST study of \citet{Whitaker08}.}

\label{table:sample}
\end{deluxetable}
\clearpage
\end{landscape}

% TABLE 2
\clearpage
\LongTables
%\begin{landscape}
\tabletypesize{\tiny}
\begin{deluxetable}{cccccccccc}
\tablecaption{Derived Quantities}
\tablewidth{0pt}
\startdata \hline\hline
Object     & $\tau$ & metallicity & age & SFR$_{\rm IR}$             & SFR$_{\rm spec}$        & SDSS class & SB09 class & SDSS M$_{*}$                  & M$_{\rm dust}$ \\
               &             &                  & Gyr  & M$_{\odot}$ yr$^{-1}$  & M$_{\odot}$ yr$^{-1}$ &                  &                   & 10$^{10}$ M$_{\odot}$     &  10$^5$ M$_{\odot}$ \\\hline
1-2874 & 0.1 & 0.008 & 9.88 & $<$0.8 & \nodata & \nodata & No emission & \nodata & $<$21 \\
2-368 & 0.3 & 0.02 (Z$_\odot$) & 9.97 & $<$0.8 & \nodata & \nodata & \nodata & \nodata & $<$21 \\
3-601 & 0.3 & 0.008 & 9.87 & $<$0.9 & \nodata & \nodata & ? & \nodata & $<$13 \\
4-1190 & 0.1 & 0.02 (Z$_\odot$) & 9.88 & $<$1.1 & \nodata & \nodata & \nodata & \nodata & $<$18 \\
4-1975 & 0.1 & 0.02 (Z$_\odot$) & 9.90 & $<$1.2 & 0.2 & -1 & \nodata & 8 & $<$19 \\
 &  &  &  &  & 
 &  \\ 
4-2713 & 0.1 & 0.008 & 9.94 & $<$0.5 & 0.3 & 2 & \nodata & 5 & $<$15 \\
5-901 & 0.1 & 0.02 (Z$_\odot$) & 9.90 & $<$0.3 & 0.2 & 4 & \nodata & 3 & $<$10 \\
5-994 & 0.1 & 0.02 (Z$_\odot$) & 9.91 & 0.2 & 0.3 & 4 & ? & 11 & 7 \\
5-1271 & 0.1 & 0.02 (Z$_\odot$) & 10.30 & 2.3 & 3.8 & 4 & \nodata & 13 & 42 \\
6-1553 & 0.3 & 0.02 (Z$_\odot$) & 9.98 & 0.0 & 0.0 & -1 & ? & 2 & $<$3 \\
 &  &  &  &  & 
 &  \\ 
6-1676 & 0.1 & 0.02 (Z$_\odot$) & 9.91 & 0.7 & 0.4 & -1 & Seyfert & 8 & 19 \\
7-471 & 0.1 & 0.02 (Z$_\odot$) & 9.92 & $<$0.8 & \nodata & \nodata & \nodata & \nodata & $<$21 \\
7-2322 & 0.1 & 0.02 (Z$_\odot$) & 9.93 & $<$0.5 & \nodata & \nodata & ? & \nodata & $<$16 \\
7-4247 & 0.1 & 0.02 (Z$_\odot$) & 9.98 & $<$0.9 & \nodata & \nodata & \nodata & \nodata & $<$14 \\
8-2119 & 0.1 & 0.02 (Z$_\odot$) & 10.26 & 1.9 & \nodata & \nodata & LINER/Seyfert & \nodata & 32 \\
 &  &  &  &  & 
 &  \\ 
9-360 & 0.1 & 0.02 (Z$_\odot$) & 9.89 & 1.0 & \nodata & \nodata & \nodata & \nodata & 17 \\
9-2105 & 0.1 & 0.02 (Z$_\odot$) & 10.30 & 1.8 & \nodata & \nodata & ? & \nodata & 31 \\
9-2966 & 0.1 & 0.02 (Z$_\odot$) & 9.96 & 0.8 & \nodata & \nodata & \nodata & \nodata & 13 \\
9-3079 & 0.1 & 0.02 (Z$_\odot$) & 9.98 & $<$0.4 & \nodata & \nodata & No emission & \nodata & $<$11 \\
10-112 & 3.0 & 0.02 (Z$_\odot$) & 10.02 & 2.3 & \nodata & \nodata & LINER & \nodata & 37 \\
 &  &  &  &  & 
 &  \\ 
10-232 & 5.0 & 0.02 (Z$_\odot$) & 9.97 & 3.4 & 1.2 & 4 & Seyfert & 7 & 45 \\
11-962 & 0.3 & 0.02 (Z$_\odot$) & 9.92 & 2.3 & 0.4 & -1 & \nodata & 28 & 34 \\
11-1278 & 0.1 & 0.02 (Z$_\odot$) & 9.87 & 0.3 & 0.2 & 4 & ? & 7 & 10 \\
11-1361 & 0.1 & 0.02 (Z$_\odot$) & 9.88 & 1.1 & 1.2 & 2 & \nodata & 16 & 19 \\
11-1732 & 0.1 & 0.02 (Z$_\odot$) & 9.84 & 0.9 & 3.7 & 4 & LINER & 14 & 15 \\
 &  &  &  &  & 
 &  \\ 
12-1734 & 0.1 & 0.02 (Z$_\odot$) & 10.30 & $<$0.6 & 0.3 & -1 & No emission & 6 & $<$19 \\
13-3813 & 0.3 & 0.02 (Z$_\odot$) & 9.93 & $<$0.3 & \nodata & \nodata & No emission & \nodata & $<$10 \\
14-1027 & 0.3 & 0.008 & 9.91 & $<$0.8 & \nodata & \nodata & \nodata & \nodata & $<$13 \\
14-1401 & 0.1 & 0.02 (Z$_\odot$) & 9.93 & $<$0.4 & \nodata & \nodata & Seyfert & \nodata & $<$14 \\
16-584 & 0.1 & 0.02 (Z$_\odot$) & 9.91 & $<$0.7 & \nodata & \nodata & \nodata & \nodata & $<$21 \\
 &  &  &  &  & 
 &  \\ 
16-650 & 0.1 & 0.02 (Z$_\odot$) & 9.88 & $<$0.7 & 0.6 & -1 & ? & 23 & $<$18 \\
16-1302 & 0.1 & 0.02 (Z$_\odot$) & 9.90 & 0.5 & 0.3 & 4 & ? & 16 & 19 \\
17-596 & 0.1 & 0.02 (Z$_\odot$) & 9.87 & $<$0.2 & \nodata & \nodata & No emission & \nodata & $<$7 \\
17-681 & 3.0 & 0.02 (Z$_\odot$) & 10.03 & 0.4 & 0.1 & -1 & No emission & 16 & 14 \\
17-2134 & 0.1 & 0.02 (Z$_\odot$) & 10.28 & 7.0 & 1.4 & 3 & ? & 13 & 104 \\
 &  &  &  &  & 
 &  \\ 
17-2819 & 5.0 & 0.02 (Z$_\odot$) & 9.93 & $<$0.7 & 0.4 & 4 & \nodata & 11 & $<$21 \\
18-476 & 0.1 & 0.008 & 9.94 & $<$0.4 & 0.1 & -1 & \nodata & 6 & $<$14 \\
18-794 & 0.1 & 0.02 (Z$_\odot$) & 9.90 & $<$0.0 & 0.1 & -1 & \nodata & 5 & 3 \\
19-2206 & 0.3 & 0.02 (Z$_\odot$) & 9.93 & $<$0.5 & \nodata & \nodata & LINER & \nodata & $<$18 \\
19-2242 & 0.1 & 0.02 (Z$_\odot$) & 9.98 & $<$0.6 & \nodata & \nodata & No emission & \nodata & $<$18 \\
 &  &  &  &  & 
 &  \\ 
20-2319 & 0.1 & 0.008 & 9.88 & $<$0.7 & 0.5 & -1 & \nodata & 8 & $<$18 \\
20-2395 & 2.0 & 0.02 (Z$_\odot$) & 10.03 & 5.7 & \nodata & \nodata & \nodata & \nodata & 79 \\
21-523 & 0.1 & 0.008 & 9.91 & $<$0.7 & 0.0 & -1 & \nodata & 6 & $<$21 \\
21-837 & 5.0 & 0.02 (Z$_\odot$) & 9.94 & $<$0.7 & 0.5 & -1 & \nodata & 11 & $<$18 \\
22-790 & 3.0 & 0.02 (Z$_\odot$) & 10.02 & 0.8 & 0.7 & -1 & No emission & 15 & 24 \\
 &  &  &  &  & 
 &  \\ 
22-2252 & 5.0 & 0.02 (Z$_\odot$) & 9.94 & 2.3 & 2.7 & 2 & \nodata & 11 & 39 \\
23-1585 & 0.1 & 0.008 & 9.93 & $<$0.3 & 0.1 & -1 & \nodata & 4 & $<$12 \\
23-2937 & 5.0 & 0.02 (Z$_\odot$) & 9.95 & 1.5 & 0.7 & 4 & \nodata & 25 & 23 \\
25-1980 & 3.0 & 0.02 (Z$_\odot$) & 10.00 & $<$0.4 & 0.6 & 4 & \nodata & 43 & $<$11 \\
25-3572 & 0.1 & 0.02 (Z$_\odot$) & 10.29 & $<$0.4 & \nodata & \nodata & \nodata & \nodata & \nodata \\
 &  &  &  &  & 
 &  \\ 
26-2558 & 0.1 & 0.008 & 9.93 & $<$0.9 & 0.3 & -1 & \nodata & 24 & $<$14 \\
26-4396 & 0.3 & 0.02 (Z$_\odot$) & 9.55 & $<$0.3 & 0.0 & -1 & \nodata & 4 & $<$12 \\
27-984 & 0.1 & 0.02 (Z$_\odot$) & 9.89 & $<$0.3 & 0.2 & 2 & \nodata & 4 & $<$12 \\
27-3444 & 0.1 & 0.02 (Z$_\odot$) & 10.00 & 1.5 & 2.5 & -1 & \nodata & 23 & 23 \\

\enddata 
\tablecomments{Subset of red galaxies from vD05 which are early type and have tidal features.  The values of $\tau$, metallicity, and age are the results of the best-fitting BC03 models to the SEDs, while SFR$_{\rm IR}$ is derived from the MIPS 24~$\micron$ point. These SED-derived quantities are described in Section \ref{sec:seds}.  The SFR$_{\rm spec}$ is derived from the SDSS DR4 spectrum, as described in \citet{Brinchmann04}.  The SDSS class also comes from \citet{Brinchmann04} and is decoded as follows:  1 = SF, 2 = Low S/N SF, 3 = composite, 4= AGN, -1 = Unclassifiable. See Section \ref{sec:bpt} for more details.  The SB09 class is based on [OII]$\lambda$3727, H$\beta$, and [OIII]$\lambda$5007 measurements made by \citet{SanchezBlazquez09}. The dust mass is derived as described in Section \ref{sec:gascontent}.}

\label{table:derived}
\end{deluxetable}
\clearpage
%\end{landscape}

\end{document}